\documentclass{chi2012}
\pdfoutput=1
\usepackage{times}
\usepackage{amsmath,amssymb,amsfonts}
\usepackage{graphicx}
\usepackage{subfigure}

\newcommand{\eat}[1]{} %TO EAT AWAY MATERIAL

\usepackage{graphics}
\usepackage{color}
\usepackage[pdftex]{hyperref}
\hypersetup{%
pdftitle={Your Title},
pdfauthor={Your Authors},
pdfkeywords={your keywords},
bookmarksnumbered,
pdfstartview={FitH},
colorlinks,
citecolor=black,
filecolor=black,
linkcolor=black,
urlcolor=black,
breaklinks=true,
}

\newcommand{\comment}[1]{}
\definecolor{Orange}{rgb}{1,0.5,0}

\begin{document}
% To make various LaTeX processors do the right thing with page size.

\setlength{\paperheight}{11in}
\setlength{\paperwidth}{8.5in}
\setlength{\pdfpageheight}{\paperheight}
\setlength{\pdfpagewidth}{\paperwidth}

% Use this command to override the default ACM copyright statement
% (e.g. for preprints). Remove for camera ready copy.
\toappear{}

\title{Swayed by Friends or by  the Crowd?}

\numberofauthors{3}
\author{
  \alignauthor Zeinab Abbassi\\
    \affaddr{Dept. of Computer Science}\\
    \affaddr{Columbia University}\\
    \email{zeinab@cs.columbia.edu}
  \alignauthor Christina Aperjis\\
    \affaddr{Social Computing Group}\\
    \affaddr{HP Labs}\\
    \email{christina.aperjis@hp.com}
     \alignauthor Bernardo A. Huberman\\
    \affaddr{Social Computing Group}\\
    \affaddr{HP Labs}\\
    \email{bernardo.huberman@hp.com}
   }
    
\maketitle

\begin{abstract}
We have conducted three empirical studies of the effects of friend recommendations and general ratings on how online users make choices. These two components of social influence were investigated through user studies on Mechanical Turk. We find that for a user deciding between two choices an additional rating star has a much larger effect than an additional friend's recommendation on the probability of selecting an item. Equally important, negative opinions from friends are more influential than positive opinions, and people exhibit more random behavior in their choices when the decision involves less cost and risk. Our results can be generalized across different demographics, implying that individuals trade off recommendations from friends and ratings in a similar fashion.

\end{abstract}


\category{K.4.4}{Computers and Society}[ Electronic Commerce]


\section{Introduction}

When making choices, people use information from a number of sources including friends, family, experts, media, and the general public. Two sources that are particularly relevant in an online setting are the opinions of friends and ratings from the general public. Friends influence choices because often their interests overlap with their friends' interests. In many cases, however, recommendations from one's friends are in stark contrast to opinions of individuals in the general public who are not one's friends.

An interesting question, which we study in this paper, is how the decision of an online user is influenced by friend recommendations and ratings from the general public, in particular when these two sources of information are in conflict with each other. This question is interesting for two reasons:  First, understanding how people trade off friends' opinions with ratings from the general public helps to determine the weight assigned by consumers to these two sources when they are uncertain about choosing one of two possible options. Second, from a normative standpoint, this information can be used when designing algorithms that display these two sources of information in order to increase the probability of a user selecting one of the options.  For example, an online social network platform that has information about how a user's friends and the general public have rated two different items can display to the user the item that she is more likely to select.  %Additionally, if an advertiser knows that a user does not trust her friends' reviews, these sources need not be shown to this user.
On the other hand, if users tend to disregard some source of information, this source need not be shown to the user.  Finally, an advertiser that wishes to make the user choose a certain item may strategically choose which pieces of information to show.

Specifically, focusing on friends and the general public as two components of social influence is important because these sources of social information are already used in a variety of algorithms and applications online. Social recommender systems take into account the actions of a user's friends and make recommendations accordingly. Social search is also gaining more attention. Google recently launched its +1 button for search results and ads in order to improve its search algorithm. If a user thinks that a search result or an ad is useful she can click on the +1 button. The +1 will be displayed along with the user's name in the search results to all her friends who subsequently search a similar query. For users who are not friends, only the number of +1's will be displayed. Facebook uses a similar approach for business pages with the intention of getting higher click-through rates. The model that we suggest can be leveraged to design better algorithms for these and other similar applications.

In addition to quantifying how an individual makes choices in the presence of social influence, we also consider how a group of users makes decisions over time when exposed to information from friends as well as ratings from the general public. While previous work has studied how groups make decisions in the presence of social influence, it has not considered the different sources of this influence. For instance, the model of Bendor et al. that considers individuals making choices takes into account one source of information, namely the number of others that have chosen one item out of two options~\cite{bhw05}. By contrast we consider how market share of the two options evolves in a setting where there are two sources of social influence, namely opinions from friends and ratings from the general public.

In particular, in this paper we asked the following questions:

\begin{itemize}

\item {\bf R1:} How much are one's choices influenced by the opinions of her friends compared to ratings from the general public? What mathematical model explains this?

\item  {\bf R2:} Do friends' negative opinions have a stronger or weaker effect than friends' positive opinions about an item?

\item {\bf R3:} Do friends' opinions have the same effect on one's decision in higher risk situations versus lower risk situations? %More specifically, is one's behavior different when monetary cost is involved compared to situations where the cost is not monetary and less serious?

\item  {\bf R4:} How does market share evolve over time in the presence of different aspects of social influence?
\end{itemize}

To answer the first three questions, we performed user studies on Mechanical Turk using positive and negative opinions from friends, as well as ratings from the general public; the latter was represented by the average number of stars. We find that the choice between two options fits a logit model. Our major findings are that (1) an additional recommendation star has a much larger effect than an additional friend's recommendation on the probability of selecting an item, (2) negative opinions from friends are more influential than positive opinions, and (3) people exhibit more random behavior in their choices when the decision involves less cost and risk.  Our results are quite general in the sense that people across different demographics trade off recommendations from friends and ratings in a similar fashion.
With respect to R4, we performed simulations and observed that the variability of outcomes increases when the variability of ratings increases, and inequality increases when individuals become more selective in their recommendations.
%To answer R4, we performed simulations in a setting with a population of individuals that make choices over time and each individual provides a rating for the item he chose.  We find that the market share of each option converges and there are multiple potential limit points.  Our main observations are that (1) the variability of limit points increases when the variability of ratings increases and (2) inequality increases when individuals become more selective in their recommendations.

\section{Related Work}
%Our work is related to a number of research areas. Many theoretical models have been proposed on how individuals make decision in the presence of social influence. There has also been some empirical studies on datasets. Another line of research that is related to the present work is on the negative-positive asymmetry. We go over these related work in the rest of this section.

\subsection{Empirical Studies on Social Influence}

A number of empirical studies have considered the effect of social influence in various contexts, including prescription drug adoption and use~\cite{IVV11}, viral and word of mouth marketing~\cite{aral2010creating}, health plans~\cite{sorensen2006social}, crime rates~\cite{glaeser1995crime}, and investment in the stock market~\cite{av10,hong2004social}.
Tucker et al. focus on how quality and popularity influence decisions on a wedding website~\cite{tucker2011does}.

Guo et al. study the role of social networks in online shopping~\cite{guo2011role}. Their study is performed on data from a leading Chinese e-commerce website, which deploys a messaging system among its users. The data contains users' purchase behavior as well as message exchanges among them; however, the authors did not have access to the content of the messages. The main finding of the paper that is related to our work is that, a message sent from a user that has purchased an item increases the probability of purchase by the user who got the message by 1\%. The causal relation is not definite, since the content of the messages is not known.

Salganik et al. study the effects of social influence over time on the popularity of songs on an artificial online music market~\cite{salganik2006experimental}. They find that social influence may lead to unpredictable and unequal outcomes.  This work is related to our study of the dynamics of market share in the presence of social influence; however, we consider two sources of social influence.  As we discuss below, the dynamics of social influence have also been analyzed theoretically.

\subsection{Theoretical Models on Dynamics under Social Influence}

Theoretical models focus mainly on the dynamics of market share in the presence of social influence: before making a decision, an individual observes what choices others have made in the past~\cite{banerjee1992simple,bikhchandani1992theory,welch1992sequential}.
%These models make different assumptions about the rationality of people ranging from full rationality~\cite{banerjee1992simple,bikhchandani1992theory,welch1992sequential} to settings where agents update their beliefs in a naive fashion~\cite{gj10}.
More related to our study are~\cite{bhw05,cms09}. Bendor et al. presented a model that considers individuals making choices for which the objective evidence of their relative value is weak~\cite{bhw05}. In that case, individuals tend to rely heavily on the prior choices of people in similar roles. They then showed that the dynamics of market shares lead to outcomes that appear to be deterministic in spite of being governed by a stochastic process. In other words, when
the objective evidence for the value of the choices is weak, any sample path of this process quickly settles down to a fraction of adopters that is not predetermined by the initial conditions: ex ante, every outcome is just as (un)likely as every other. In~\cite{cms09}, the authors define the decision function as a stochastic function of the social influence (in terms of market share) and the inherent quality of the product. Their main result is that market shares converge to an equilibrium.
These processes only take into account one source of social information, namely the number of others which have chosen one item over the other.  By contrast we consider how market share evolves in a setting with two sources of social influence, namely opinions from friends and ratings from the general public.

\subsection{Sources of Social Influence}
Most previous work models choosing between items as a function of two factors: the individual's own judgment and social influence. However, different groups of people have different levels of influence on one's decision. For example, when one wants to choose a movie to watch, some people rely more on what the general ratings for the movie are and some others more on what their friends who have already watched the movie say about it. Currently, with the availability of social network data, in many online settings, recommendations from friends are available in addition to ratings and reviews from other people. Due to this fact, in this work we focus on these two sources of social influence. 

\section{Method}

In this section we describe the experimental design and report some statistics about the data we collected. Our goal is to study how people trade off information from friends and the general public when choosing between two items.
Moreover, our experiments allow us to compare a setting where the information from friends consists of positive recommendations to a setting where the information from friends consists of negative opinions.

Furthermore, we compare people's choices with respect to two types of decisions: one that involves a monetary cost (booking a hotel) and a low risk decision that involves no monetary cost (watching a movie trailer).
We chose booking hotels because the user cannot go and check it out before deciding and should rely on the information she gets from others.  Similarly, a user may not have any information about a movie trailer before she watches it.  We can think of the setting with the movie trailers as a less serious decision, since it involves less cost (just a couple of minutes of one's time) and risk.  Users often make choices of this type online, e.g., when watching Youtube videos, clicking on a link or ad, etc.

In total, we conducted three user studies: booking a hotel with positive recommendations from friends (Study 1), booking a hotel with negative opinions from friends (Study 2) and watching a movie trailer with positive recommendations from friends (Study 3).

\eat{
\begin{center}
\begin{figure}

\includegraphics[width=3.33in, height=2.5in]{sample-survey.pdf}
\label{sample-survey}
\end{figure}
\end{center}

\begin{center}
\begin{figure}

\includegraphics[width=3.33in, height=1.22in]{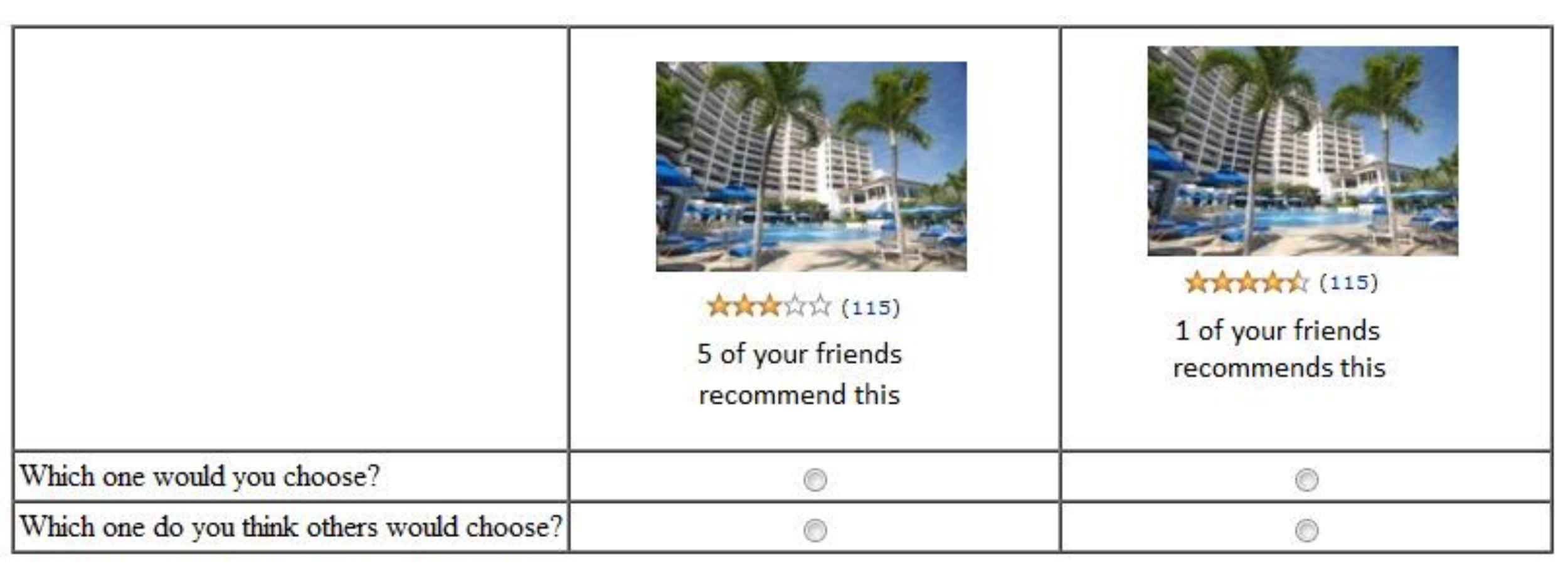}
\label{sample-survey}
\end{figure}
\end{center}
}

\noindent{\bf Online survey}\\
To collect the data we conducted the three studies in the form of surveys on Amazon's Mechanical Turk during July and August 2011. Amazon's Mechanical Turk (MTurk) is a crowdsourcing online marketplace where \textit{requesters} use human intelligence of \textit{workers} to perform certain tasks, also known as HITs (Human Intelligence Tasks). Workers browse among existing tasks and complete them for a monetary payment set by the requester~\cite{mason2010conducting}.

Once a worker completes the task, the requester can decide whether to approve it.  In particular, if the requester believes that the worker completed the task randomly, he can reject her work.  In that case, the worker does not get paid for the particular task and her approval rate is decreased.
For our studies, we only hired workers that had approval rates of over 95\%, that is, workers who had performed well in the past.

We asked each worker to put herself in the following hypothetical situation: she is about to book a hotel (resp. watch a movie trailer) an on e-commerce site (resp. online), and among the options, she has come down to two between which she is indifferent.  The website has an underlying social network of friends (or it runs on top of an online social network).  For each of the two options, we provide the following information:
\begin{enumerate}
\item [(i)] the overall rating (in terms of stars on the scale of 1 to 5) based on ratings from a large number of previous customers (resp. users) in the case of selecting which hotel to book (resp. which movie trailer to watch)
\item [(ii)] the number of friends who recommend (resp. have negative opinions about) the option in the case of positive (resp. negative) recommendations
\end{enumerate}
For each question, the option that has more stars is the one that is less recommended by friends; that is, we did not use a pair of options where one clearly dominated the other.

\begin{center}
\begin{figure}
\includegraphics[width=3.33in, height=1.22in]{positive-survey.pdf}
\includegraphics[width=3.33in, height=1.22in]{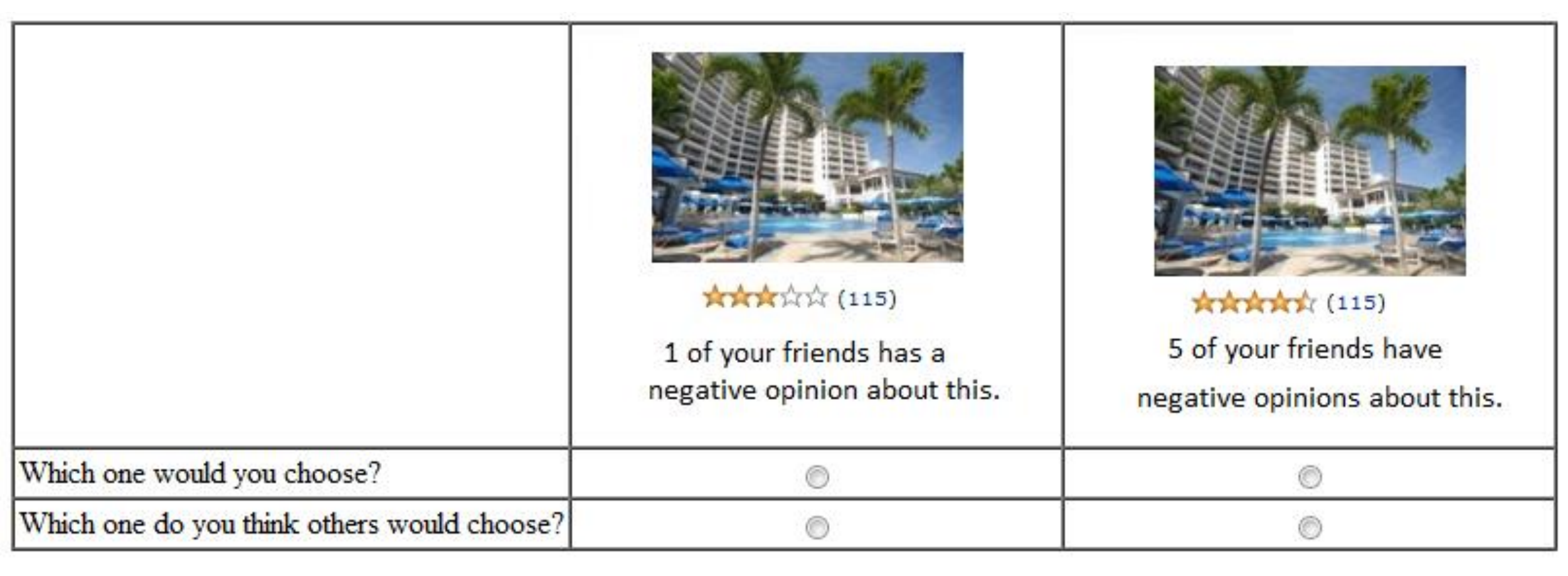}
\includegraphics[width=3.33in, height=1.22in]{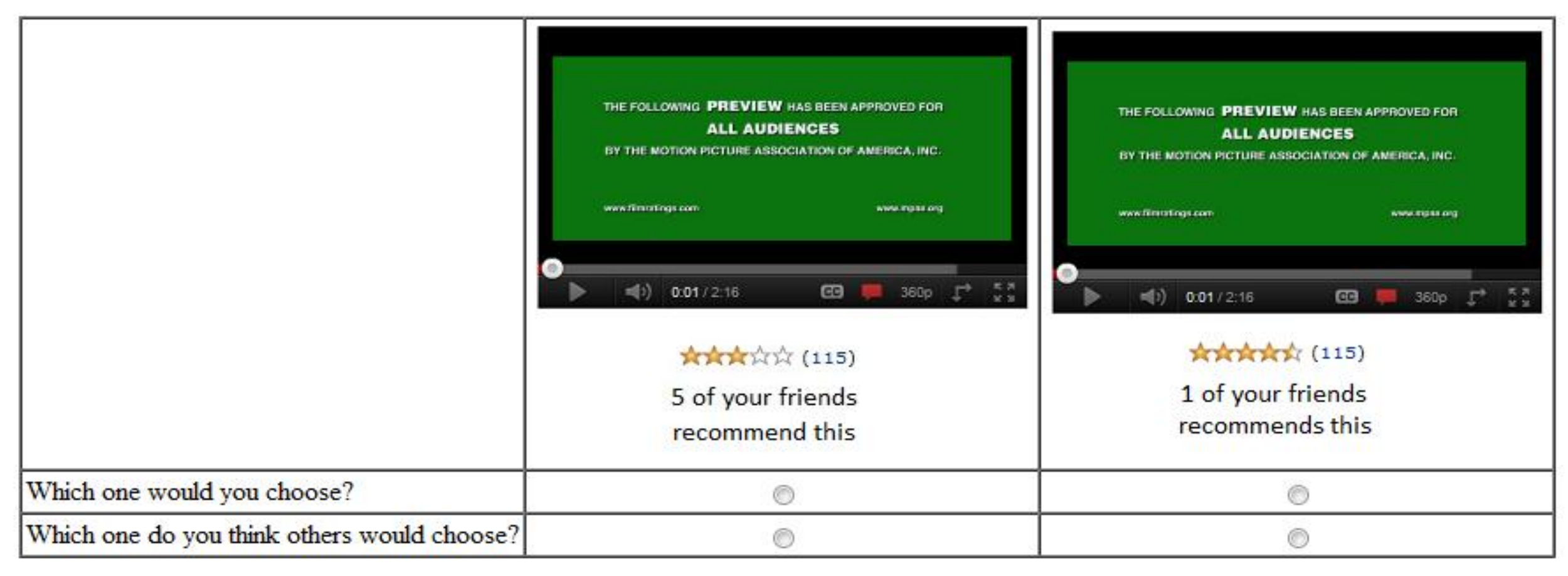}
\caption{Sample questions for Studies 1, 2, and 3 (top to bottom).}
\label{sample-survey}
\end{figure}
\end{center}

A sample question from each survey is shown in Figure~\ref{sample-survey}.
Each question consists of two parts: 1) \textit {``Which one would you choose?"} and 2) \textit{``Which one do you think others would choose?"}   The answer to the first question provides information on how a particular worker trades off information from friends and the general public when making her choice.  We only use the data from this question for our analysis.

Even though we do not use the answers to the second question for our analysis, we included it in the survey for two reasons.  First, by asking the second question the subjects felt that the survey's goal was to study how people think their decisions are similar to the others, weakening the reactivity effect~\cite{heppner2008research}. Second, we offered three bonus payments (\$5 or \$10 each, which is 50 or 100 times the amount we paid for each HIT) to the three workers whose answers to the second question was closest to the others' answers to the first question in order to deter workers from answering randomly.

{\bf \noindent Validity check.}
Apart from using the bonus to incentivize workers to put some thought when answering the survey, we incorporated two ``validity check'' questions in the survey.  If a worker did not answer these two questions correctly, we did not include any of her responses in our analysis (and rejected her work).
In the first such question one option clearly dominated the other in terms of both the number of stars and friends' recommendations (one option had only one star and 10 negative recommendations from friends, whereas the other had more stars and 10 positive recommendations from friends).
The second test question was a repeated question with the order of choices reversed and also with no graphics. Responses with inconsistencies in the answers to these two questions were discarded and not considered in the analysis. We ran the experiments until we got 350 valid responses for each study. Overall, we rejected 33\% of the responses across all 3 studies because they were invalid. The average completion time for each valid HIT was 174.8 seconds while the average completion time for the invalid HITS was 153.3; this suggests that the workers that were rejected had not taken the task as seriously as the rest of the workers.

\begin{center}
\begin{figure}
\includegraphics[width=3.33in, height=2.5in]{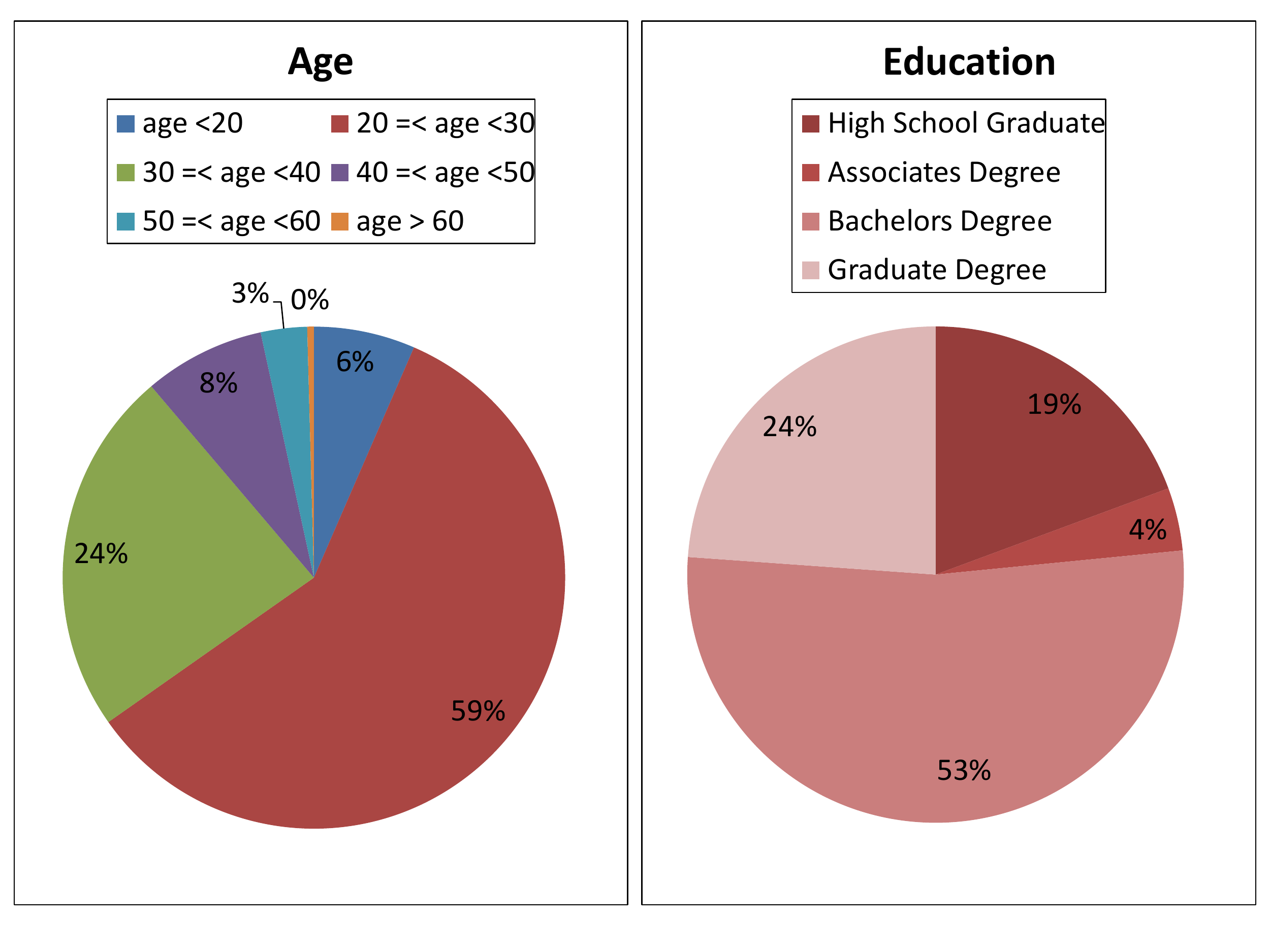}
\caption{Demographics}
\label{plots}
\end{figure}
\end{center}

In addition to the ``validity check'' questions, each study consisted of 8 questions (with the format of Figure \ref{sample-survey}) which we use in our analysis and 3 demographics questions asking about the gender, the age and the education level of the respondent. Overall, 36\% of the approved respondents, were female. Other demographic information for the approved respondents according to the self-reporting of the workers are plotted in Figure~\ref{plots}.

\subsection{Model Specification}
\label{modelspec}
For each question, there are two options that the worker can select form, which we refer to as option $1$ and option $2$. We denote the number of stars $S_i$ and the number of friend recommendations by $F_i$, for $i=1,2$.
To predict the probability that option 1 is selected, we conduct a logistic regression on our dataset of choices with the difference in the number of stars (i.e., $S_1-S_2$) and friends (i.e., $F_1-F_2$) for each question as the predictor variables. %For our choice variable we gave the first choice (option $A$) a value of 1 and the second choice (option $B$) a value of zero.
In particular,
\[Pr[\text{choose option 1}] = \mbox{logit}(\alpha_s \cdot (S_1-S_2) + \alpha_f \cdot (F_1-F_2)).\]

We note that a number of other empirical studies also use the logit choice function to model social influence~\cite{paez2008discrete, glaeser1995crime, sorensen2006social}.

%Logistic regression (logit) is a standard way to model dichotomous outcome variables. In logistic regression, the log odds of the outcome is modeled as a linear combination of the predictor variables.

%Basically, the formula is: $$Pr[A] = {{e^{\alpha_s S_A + \alpha_f F_A}}\over{e^{\alpha_s S_A + \alpha_f F_A}+e^{\alpha_s S_B + \alpha_f F_B}}} ={{1}\over{1+e^{\alpha_s(S_B-S_A) + \alpha_f(F_B-F_A)}}} $$

\section{Results}
\label{results}

In this section, we report the results from each survey separately.\\ \\

%For each study 350 valid results were collected. In this section we report the results of our two studies: positive recommendations versus anonymous ratings and negative recommendations versus anonymous ratings.

\subsection{Study 1: Positive Opinions for Hotels}
\subsubsection{Model 1}
As mentioned in the previous section, we examine the results by fitting a logistic regression model. We first only considered the difference in the number of stars and friends as predictor variables.
The estimated coefficients along with other parameters are depicted in Table~\ref{pos}, and as can be seen both are statistically significant.  Observe that both coefficients are positive; this is intuitive, since more stars (resp. more positive recommendations) indicate that the option is better and thus the worker is more likely to select it.
Finally, the pseudo~-$R^2$ for this model\footnote{We computed Efron's pseudo~-$R^2$ which is defined as follows:
$$R^2 = 1 - {{\sum_{i=1}^N (y_i - \hat{\pi_i})^2}\over{\sum_{i=1}^N (y_i - \bar{y})^2}}$$ where, $N$ is the number of observations in the model, $y$ is the dependent variable, $\bar{y}$ is the mean of the $y$ values, and $\hat{\pi}$ is the probabilities predicted by the logit model. The numerator of the ratio is the sum of the squared differences between the actual $y$ values and the predicted $\pi$ probabilities.  The denominator of the ratio is the sum of squared differences between the actual $y$ values and their mean~\cite{HHH07}.
}
is 0.95, indicating that the fit is very good.

\begin{table}
\begin{center}
\caption{Positive}
\label{pos}
\begin{tabular}{c|c c}
  \hline
  Predictor & {Estimated Coefficients} & {z-value} \\
  \hline
$\alpha_f$ & 0.20471$^{\small{***}}$ \small{(0.027)}& 7.597\\

$\alpha_s$ & 0.73549$^{\small{***}}$ \small{(0.050)} & 14.307 \\
  \hline
 \end{tabular}\\
\begin{tabular}{l}
\small {Note: Standard errors are shown in parentheses.}\\
\small{Signif. codes:  0 '***' 0.001 '**' 0.01 '*' 0.05 '.' 0.1 ' ' 1}\\
\small {Pseudo-$R^2$ = 0.95}\\
%\small {Log-Likelihood = -1717.576}\\
\hline
\end{tabular}
\end{center}
\end{table}

{\bf Interpretation of the coefficients} We interpret the coefficients for our model in terms of marginal effects on the odds ratio. The odds ratio measures the probability that the dependent variable is equal to 1 relative to the probability that it is equal to zero.
For the logit model, the log odds of the outcome is modeled as a linear combination of the predictor variables; therefore, the odds ratio of a coefficient is equal to $\exp$(coefficient). Since $\alpha_s = 0.735$, we conclude that a unit increase in $S_1 - S_2$, multiplies the initial odds ratio by $\exp(0.735) = 2.07$. In other words, the relative probability of choosing option $1$ increases by 107\%. For the friends predictor variable, the odds ratio is equal to $\exp$(0.204) = 1.22, meaning that, the relative probability of selecting option $1$ increase by 22\% if $F_1 - F_2$ increases by one unit.

To further assess the predictive power of the model, we performed cross validation. We left out one question at a time and estimated the coefficients using the remaining questions. Then, we predicted the probabilities for the question that was left out. The predicted values were very close in all cases with absolute mean difference of 0.021. The actual values and their differences can be found in Table~\ref{crossvalidation}.

\begin{table}
\begin{center}
\caption{Cross validation for study 1}
\label{crossvalidation}
\begin{tabular}{c c c c}
\hline
Left out question & Actual & Predicted & $|$Difference$|$\\
\hline
Q1& 0.54 &0.53 & 0.01 \\
Q2 & 0.58 & 0.55 & 0.03 \\
Q3 & 0.71 & 0.62 & 0.09 \\
Q4 & 0.74 & 0.74 & 0.00 \\
Q5 & 0.77 & 0.77 & 0.00\\
Q6 & 0.74 & 0.72 & 0.02\\
Q7 & 0.82 & 0.83 & 0.01\\
Q8 & 0.54 & 0.53 & 0.01\\
\hline
\end{tabular}
\end{center}
\end{table}

Finally, we used one of the questions of this survey twice in Amazon's Mechanical Turk (in two separate HITS) in order to see whether workers would react to the question in similar ways.  We found that the percentage of workers that chose the first option of the question was similar in both cases (26\% versus 24\%), further validating our approach.

%In the next stage we included other predictor variables such as gender, education and age.
\subsubsection{Model 1'}
In Model 1', we included all self reported demographic information as predictor variables in addition to the stars and friend recommendation variables. This information includes: gender, age (in five 10-year brackets from 20 to 60 years old: called age1-5), and education level (high school, associates degree, bachelors degree, and  graduate degree: called edu1-3). More specifically, we coded the following variables as dummy variables. The estimated coefficients and other information is shown in the second column of Table~\ref{demographics}. As can be seen in Table~\ref{demographics}, these extra coefficients are not statistically significant. This suggests that people in different demographics trade off ratings from the public and friend recommendations similarly.\\ \\ \\

\eat{
gender as one of the predictor variables. In particular, we introduced a dummy variable $D$, which was set to 1 for female respondents and 0 for male respondents, in order to examine the effect of gender on decision making:
\begin{align*}
Pr[\text{choose 1}] =\mbox{logit}&(\alpha_s \cdot (S_1-S_2) + \gamma_s \cdot D (S_1-S_2) \\
                           &+ \alpha_f \cdot (F_1-F_2) + \gamma_f \cdot D (F_1-F_2))
\end{align*}
In this model, $\alpha_s$ and $\alpha_f$ is the coefficient for the number of stars and friends for male respondents respectively. On the other hand, the coefficient for the number of stars and friends for female respondents is $\alpha_s +\gamma_s$ and $\alpha_f + \gamma_f$ respectively. The estimated coefficients for this model are reported in Table~\ref{gender-pos}.
As can be seen in Table~\ref{gender-pos}, $\gamma_s$ and $\gamma_f$ are not statistically significant at the 1\% level.
This suggests that both men and women trade off information from friends and aggregate ratings from the general public similarly.
}
\begin{table}
\begin{center}
\caption{Stduies 1 and 2: demographic information included}
\label{demographics}
\begin{tabular}{c|c c}
  \hline
  Predictor & {Study 1} & {Study 2} \\
  \hline
$\alpha_f$ & 0.263$^{\small{***}}$ \small{(0.19)}&  0.35$^{\small{***}}$ \small{(0.07)}\\

$\alpha_s$ & 0.793$^{\small{***}}$ \small{(0.08)} & 0.66$^{\small{***}}$ \small{(0.12)} \\

gender$_f$ & -0.006$^{\small{}}$ \small{(0.07)}& -0.02$^{\small{}}$ \small{(0.06)}\\

gender$_s$ & 0.31$^{\small{.}}$ \small{(0.16)}& -0.13$^{\small{}}$ \small{(0.11)}\\

edu1$_s$ & 0.11$^{\small{}}$ \small{(0.16)} &  -0.01$^{\small{}}$ \small{(0.11)}\\

edu1$_f$ & 0.04$^{\small{}}$ \small{(0.08)}& -0.02$^{\small{}}$ \small{(0.07)}\\

edu2$_s$ & -0.11$^{\small{}}$ \small{(0.46)} &  $0.16^{\small{}}$ \small{(0.40)}\\

edu2$_f$ & -0.16$^{\small{}}$ \small{(0.23)}& $0.18^{\small{}}$ \small{(0.63)}\\

edu3$_s$ &-0.05$^{\small{}}$ \small{(0.16)} &  -0.03$^{\small{}}$ \small{(-3.1)}\\

edu3$_f$ & -0.02$^{\small{}}$ \small{(0.08)}& -0.24$^{\small{}}$ \small{(-0.31)}\\

age1$_s$ & 0.061$^{\small{}}$ \small{(0.39)} & -0.12$^{\small{}}$ \small{(0.15)}\\

age1$_f$ & 0.02$^{\small{}}$ \small{(0.21)}& -0.14$^{\small{}}$ \small{(0.11)}\\

age2$_s$ & 0.06$^{\small{}}$ \small{(0.16)} & 0.17$^{\small{}}$ \small{(0.12)}\\

age2$_f$ & 0.11$^{\small{}}$ \small{(0.08)}& 0.16$^{\small{}}$ \small{(0.08)}\\

age3$_s$ & -0.08$^{\small{}}$ \small{(0.34)} & -0.05$^{\small{}}$ \small{(0.16)}\\

age3$_f$ & -0.02$^{\small{}}$ \small{(0.18)}& -0.13$^{\small{}}$ \small{(0.10)}\\

age4$_s$ &-0.05$^{\small{}}$ \small{(0.16)} & 0.10$^{\small{}}$ \small{(0.11)}\\

age4$_f$ & -0.17$^{\small{}}$ \small{(0.08)}& 0.04$^{\small{}}$ \small{(0.07)}\\

age5$_s$ & 0.16$^{\small{}}$ \small{(0.40)} & 0.05$^{\small{}}$ \small{(0.24)} \\

age5$_f$ & 0.03$^{\small{}}$ \small{(0.22)}& 0.15$^{\small{}}$ \small{(0.17)}\\

\hline
 \end{tabular}\\
\begin{tabular}{l}
\small {Note: Standard errors are shown in parentheses.}\\
\small{Signif. codes:  0 '***' 0.001 '**' 0.01 '*' 0.05 '.' 0.1 ' ' 1}\\
%\small {Pseudo-$R^2$ = }\\
%\small {Log-Likelihood = }\\
\hline
\end{tabular}
\end{center}
\end{table}

\eat{
We also included gender, age, education as predictor variables (one at a time) and ran the logistic regression on our data. None of these variables were statistically significant. implying that the stars and friends describe the model very well.
}%eat

\subsection{Study 2: Negative Opinions for Hotels}
\subsubsection{Model 2}
In the previous subsection, we examined how positive recommendations from friends and the number of stars influence users' choices. In this section we look at negative opinions from friends --- instead of positive recommendations. In particular, each option is characterized by the number of stars (based on information from the general public) as well as the number of friends who have negative opinions about it. We examine the data by fitting a logit model. The estimated coefficients are reported in Table~\ref{neg}. As can be seen in the table both variables are statistically significant and the pseudo-R$^2$ measure for this model is 0.95 which implies that the model is a good fit.  Moreover, as we would expect, the friends coefficient is negative in this case, as more negative opinions from friends decrease the probability that the worker selects an option.

\begin{table}
\begin{center}
\caption{Negative}
\label{neg}
\begin{tabular}{c|c c}
  \hline
  Predictor & {Estimated Coefficients} & {z-value} \\
  \hline
$\alpha_f$ & -0.281$^{\small{***}}$ \small{(0.030)}& 9.378\\

$\alpha_s$ & 0.503$^{\small{***}}$ \small{(0.050)} & 10.018 \\
  \hline
 \end{tabular}\\
\begin{tabular}{l}
\small {Note: Standard errors are shown in parentheses.}\\
\small{Signif. codes:  0 '***' 0.001 '**' 0.01 '*' 0.05 '.' 0.1 ' ' 1}\\
\small {Pseudo-$R^2$ = 0.95}\\
%\small {Log-Likelihood = }\\
\hline
\end{tabular}
\end{center}
\end{table}

{\bf Interpretation of the coefficients} Similarly to Study 1, we interpret the coefficients for our model in terms of marginal effects on the odds ratio. For the present model (negative recommendations), the fact that  $\alpha_s = 0.503$ means that one unit increase in $S_1 - S_2$, multiplies the initial odds ratio by $\exp$(0.503) = 1.65. In other words, the relative probability of choosing option $1$ increases by 65\%. For the friends predictor variable, the odds ratio is equal to $\exp(-0.281) =0.75$, which means that the relative probability of selecting option 1 decreasing by 25\%.  Equivalently, the relative probability of selecting option $1$ when $F_1 - F_2$ decreases by one unit is $(\exp(0.281) -1) \approx 32\%$. For this study we did cross validation as well to test the predictive power of our model. The results are shown in Table~\ref{n-crossvalidation}. The predicted and actual values are very close (mean absolute difference = 0.231).

\begin{table}
\begin{center}
\caption{Cross validation for study 2}
\label{n-crossvalidation}
\begin{tabular}{c c c c}
\hline
Left out question & Actual & Predicted & $|$Difference$|$\\
\hline
Q1& 0.30 &0.25 & 0.05 \\
Q2 & 0.39 & 0.41 & 0.02 \\
Q3 & 0.43 & 0.44 & 0.01 \\
Q4 & 0.54 & 0.53 & 0.01 \\
Q5 & 0.58 & 0.60 & 0.02\\
Q6 & 0.38 & 0.39 & 0.01\\
Q7 & 0.40 & 0.42 & 0.02\\
Q8 & 0.45 & 0.50 & 0.05\\
\hline
\end{tabular}
\end{center}
\end{table}

\subsubsection{Model 2'}
Similarly to Model 1', in this model we include all variables: stars, friends' opinions, and demographics information in the model. The results are shown in the last column of Table~\ref{demographics}. As for Model 1', the estimated demographic coefficients are not statistically significant, meaning that the addition of demographic information does not improve the predictive power (compared to Model 2). In other words, individuals choose between options in these situations similarly across all demographics.

\subsection{Study 3: Positive Opinions for Movie Trailers}
\subsubsection{Model 3}
Our third study considers the effect of positive recommendations from friends in a low risk decision: choosing which movie trailer to watch.  We perform a logistic regression and report the estimated coefficients in Table~\ref{video}. The estimated coefficients are statistically significant; however, in this case pseudo-$R^2$ is 0.61 which is lower than the pseudo-$R^2$'s for previous models Models 1 and 2 (0.95). The coefficients for stars and friends are $\alpha_s =$ 0.349 and  $\alpha_f =$ 0.167.  As for Model 1, both coefficients are positive, since people are more likely to select an option if it has more stars and/or more positive recommendations from friends.  By computing the odds ratios, we conclude that an additional star increases the probability of selecting that option by 41\%, whereas an additional friend recommendation increases the probability by 18\%.

\begin{table}
\begin{center}
\caption{Less serious decision}
\label{video}
\begin{tabular}{c|c c}
  \hline
  Predictor & {Estimated Coefficients} & {z-value} \\
  \hline
$\alpha_f$ & 0.167$^{\small{***}}$ \small{(0.049)}& 7.101\\

$\alpha_s$ & 0.349$^{\small{***}}$ \small{(0.027)} & 6.014 \\
  \hline
 \end{tabular}\\
\begin{tabular}{l}
\small {Note: Standard errors are shown in parentheses.}\\
\small{Signif. codes:  0 '***' 0.001 '**' 0.01 '*' 0.05 '.' 0.1 ' ' 1}\\
\small {Pseudo-$R^2$ = 0.61}\\
%\small {Log-Likelihood = }\\
\hline
\end{tabular}
\end{center}
\end{table}

\section{Discussion}
\label{sec:discussion}

This paper studies how positive and negative opinions from friends affect our decisions compared to ratings from the crowd for different types of decisions.  Our three user studies result in some interesting conceptual findings about the tradeoff between these two types of social influence.

First, we see that an additional star has a much larger effect than an additional friend recommendation on the probability of selecting an item.  In particular, in all three studies one more star increases the probability of selecting that option more than one more (resp. less) friend in the case of positive (resp. negative) recommendations.  Equivalently, the odds ratio of the stars' coefficient is larger than the odds ratio of the friends' coefficient (2.07 versus 1.22, 1.65 versus 1.32 and 1.41 versus 1.18 for studies 1, 2 and 3 respectively).  This does not mean that the number of friends' positive or negative recommendations does not influence decisions; on the contrary, an additional recommendation (resp. one less negative opinion) from friends changes the probability by at least 18\% across all three studies.  The fact that an additional star has a larger effect that an additional friend opinion is reasonable if we consider that the number of stars is bounded between 1 and 5, whereas the number of friends' recommendations may take values from a larger range.

%From Study 1~\ref{study1} for positive recommendations, we find that one more star difference approximately doubles the odds of choosing the option with more stars; more precisely, the relative increase in the probability of choosing that option is 107\%. On the other hand, one more friend multiplies the odds of choosing the option with more friends by 1.22, and thus the relative probability increase is 22\%.
%One point that should not be ignored is that the number of stars is bound to five but the number of friends can be unlimited.

Second, negative opinions from friends are more influential on one's decision than positive opinions.
We can see this by comparing the odds ratios of Study 1 and Study 2, in which the number of positive and negative friends' opinions are shown respectively: the odds ratio for the friends variable is higher in Study 2 (1.32 versus 1.22), whereas the odds ratio for the stars variable is higher in Study 1 (2.07 versus 1.65).  In other words, one less negative opinion from a friend has a larger effect than one more positive opinion, whereas one more star increases the odds of an option being chosen less in the case that negative opinions from friends are present.
%From the second study (\ref{study2}), which looks at negative opinions from friends versus ratings, we find out that the odds ratios and therefore relative probability increases are different. In this case, the odds ratio is lower for the stars variable: 1.65 or put differently, the relative probability increase in case of one more star while all other parameters are kept the same is 65\%. This is lower than study 1, meaning that one added star increases the odds of an option being chosen less in the case were negative opinions from friends are present.
%Conversely,  the odds ratio for the friends variable is higher in Study 2: 1.32 (i.e. the relative probability increase in case of one added friend is 32\%.) %say something about robustness of this conclusion here.
Such an asymmetry between the effect of negative and positive actions and opinions have been studied in the social psychology literature~\cite{baumeister2001bad, peeters1990positive, taylor1991asymmetrical}. The \textit{positive-negative asymmetry effect} has been observed in many domains such as impression formation~\cite{Anderson65}, information-integration paradigm~\cite{anderson81} and prospect theory for decision making under risk~\cite{kt79}. The finding in all the above cited work is that negativity has stronger effects than  equally intense positivity. Our results confirm this finding in online settings.

\eat{
[to do: cite previous work on positive vs. negative opinions and mention that our findings agree with that]
}
%This might be explained with the fact that people try to avoid risks and one of the reasons they ask others about opinions is in fact this aspect [does this make sense?]

Third, people exhibit more random behavior when the decision involves less cost and less risk.   We can see this by comparing the results from Study 1 and Study 3, where the decisions are ``which hotel to book'' and ``which movie trailer to watch'' respectively.  Booking a hotel clearly involves a monetary cost and some risk, whereas the worse thing that can happen with a movie trailer is to waste a couple of minutes of one's time.  The odds ratios are lower in Study 3 than Study 1 (1.18 versus 1.22 for friends, and 1.41 versus 2.07 for stars).  This implies that one added star or friend has a smaller influence on one's decision in the case of movie trailers.  Moreover, the fraction of respondents choosing either option is closer to half compared to the hotel booking surveys. This implies that the choices were more random in this case, which may be explained by the fact that choosing which movie trailer to watch is a less important/serious decision than booking a hotel.

Forth, for all of our user studies, we find out that the demographic variables (gender, age, and education level) do not significantly impact the choice that is made, implying that people across different demographics trade off recommendations from friends and ratings from the crowd in a similar way.

In addition to the aforementioned conceptual findings, we have estimated a mathematical model that describes how users trade off friends' opinions and recommendation from the crowd in a variety of settings.

\subsection{Practical Implications}
%{\bf Practical Implications}
%Studying how friends' opinions influence users' actions is important due to the availability and popularity of social network sites.
Our studies offer insights that can be useful in various online domains such as recommender systems, social search results ranking, online advertisement placement, online social network newsfeed rankings, and social shopping websites. In these applications when both friends' recommendations and ratings from the general public are available, our estimated model can help the platform determine which option to display or what ranking to display the options for a given user.

As an example, consider a specific user that is searching for a hotel on a booking website.  There are two hotels that match the user's search criteria, hotel $A$ and hotel $B$.  Assume that hotel $A$ has 3 stars from customer ratings and 4 (positive) recommendations from the user's friends, and hotel $B$ has 4 stars but only 2 (positive) recommendations from friends.  According to the results of Study 1, the user is more likely to prefer hotel B (if everything else is equal). Thus, if the booking website does not have any additional information about the user's preferences, it should recommend hotel $B$ to the user, or equivalently rank hotel $B$ higher than hotel $A$ if it provides personalized ranking of hotels to the user.
Such personalization benefits both the user and the booking website by improving user experience and increase the chances that the user books a hotel through the website..

The same ideas can be applied to recommender systems based on collaborative filtering and in particular social recommender systems. Social recommender systems, leverage users' friends' actions to determine which items to recommend. The choice and ranking of the items can be obtained using our model. The same is true for social search.
Finally, a marketer that wishes to maximize the probability that a user selects a given item may be able to strategically select what information to show to the user.

{\bf Limitations.}
In our studies, users could only see the {\em number} of friends that had positive or negative opinions about an item --- and not the names of the corresponding friends.   We focused on the number of friends, because in this way we can get more general qualitative results.  Moreover, given that people tend to have a large number of friends in online social networks, showing the number of friends (instead of specific names) may be a good way to avoid privacy concerns. Nevertheless, we note that opinions from specific friends could have a different effect than the number of friends and that this can be an interesting future direction. 
\eat{
%[to do: elaborate on these with some examples.]
I\begin{itemize}
\item Recommender systems (in particular social recommender systems) rely on a user's friends' actions and preferences to recommend items. Many online recommender systems such as Netflix for movies show ratings from the general public,
These online applications have the opportunity to utilize social networks of their users and by displaying
\item Social shopping websites such as Blippy are gaining more popularity.
\item E-commerece websites: add a social component
\item Achieving prefect stream or newsfeed on online social networking sites such as Facebook and Google+: In the
\item Social search: search result ranking.
\item Online advertisement ranking:

\end{itemize}
}

%+discuss potential concavity 

\section{Dynamics of Market Share}
\label{dynamics}

Thus far we have studied how an individual chooses between two items when the information she has consists of the opinions of her friends and ratings from the general public.  This means that the information that is available to the individual when she makes her decision depends on the decisions that her friends and other consumers/users have made in the past.  In this section we consider a setting with a population of individuals that make decisions one at a time and the information available to someone at time $t$ depends on what others chose (and what they thought about what they chose) before time $t$.  Given this process, we study the dynamics of market share.

Previous literature has studied the effects of social influence in dynamic settings~\cite{banerjee1992simple,bikhchandani1992theory,welch1992sequential,gj10,bhw05, cms09}.  Part of this literature focuses on whether market share converges and the properties of its limit points.  The decision of an individual at time $t$ is usually assumed to depend on the market share at that time.  In contrast to previous literature, we consider two types of social influence: friends and the general public.  Moreover, we consider a setting with ratings and incorporate the individuals' rating decisions in the process.  

In order to focus on the effect of friends and the general public on one's decision (which is the central theme of this paper), we assume that an individual is otherwise indifferent between option 1 and 2 before making her decision.  In particular, we assume that she has no additional private information that affects her decision.  However, we note that it is straightforward to incorporate the effect of private information in our simulations.

Given a social network and two options (say $1$ and $2$), we consider the following process.  
\begin{itemize}
\item {\em Initialization:} At $t=0$ two individuals are randomly chosen.  One of them is assigned option 1 and the other option 2.  %These individuals rate the items with $S_1^0$ and $S_2^0$ stars respectively.
\item At time $t$, an individual that has not made a choice is randomly selected.  
    \begin{itemize}
    \item  {\em Selection:} The chosen individual selects option 1 with probability
    \[\text{logit}(\alpha_s \cdot (S_1(t) - S_2(t)) + \alpha_f \cdot (F_1(t) - F_2(t)),\]
    where $S_i(t)$ is the average number of stars for option $i$ up to time $t$, and $F_i(t)$ is the number of friends of the individual that have previously chosen option $i$ and recommend it.
    \item  {\em Rating:} The individual provides a rating for the option she chose.  If she chose option $i$, her rating is drawn from a Gaussian distribution with mean $S_i(t)$ and standard deviation $\sigma$.
    \item {\em Recommendation:} The individual recommends the item she chose to her friends if her rating exceeds some threshold $\theta$.
    \end{itemize}
\end{itemize}
For our simulations, we set $\alpha_s = 0.73$, $\alpha_f = 0.22$ (which are the estimated coefficients from Study 1),  
%$S_1^0 = $ and $S_2^0 = $, 
but note that other choices for these parameters give qualitatively similar results.  We study how the market share of product 1 evolves over time for various values of the standard deviation $\sigma$ and the threshold $\theta$.

We note that when the threshold $\theta$ is equal to zero, an individual always recommends the option she selects to her friends.  This is similar to models in prior literature (e.g.,~\cite{bhw05, cms09}), where an individual's decision is affected by how many others already bought a product (e.g., its market share) and not by the opinions of those people about the product.

\begin{center}
\begin{figure}
\includegraphics[width=3.22in, height=2.25in]{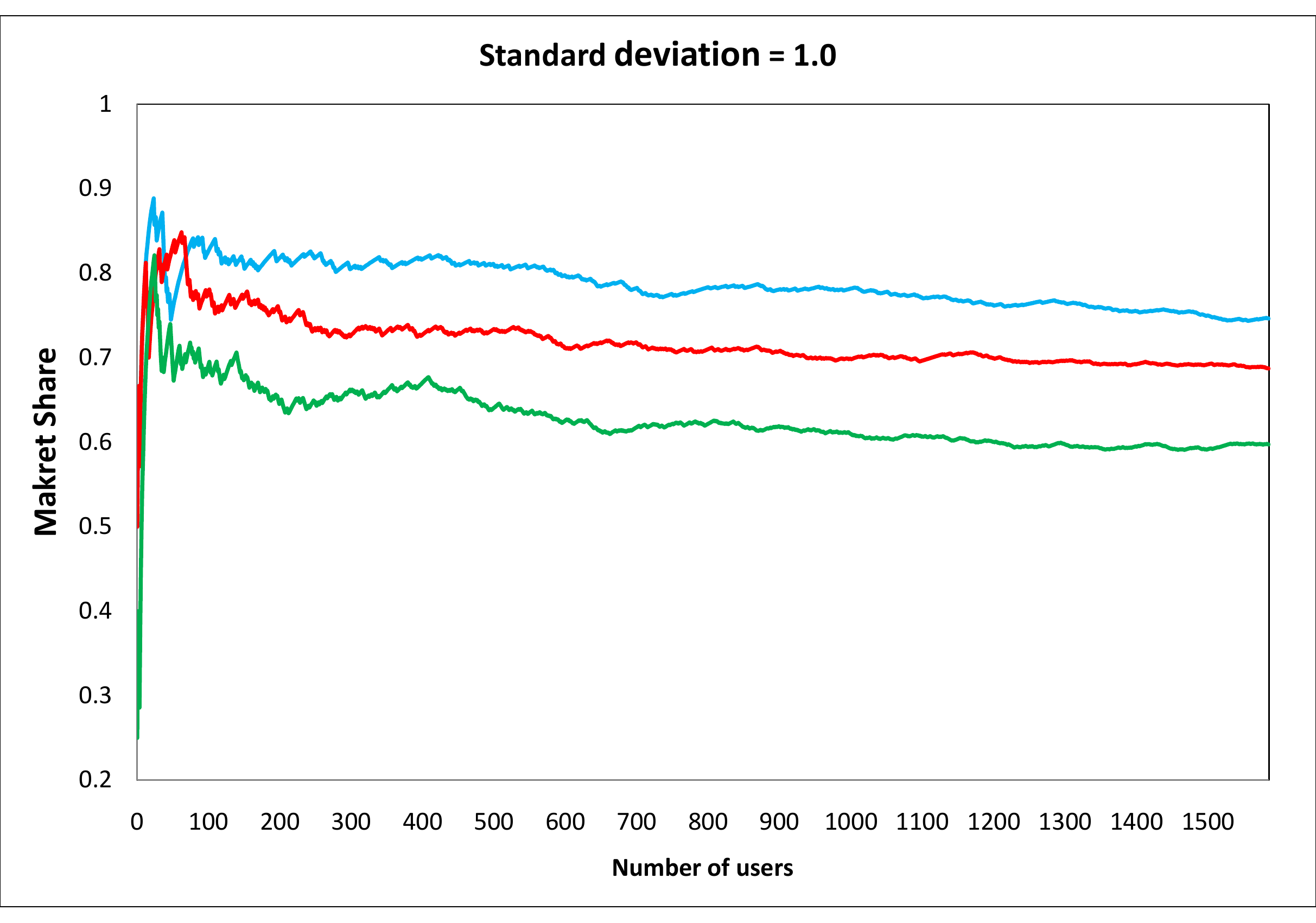}
\caption{3 sample paths of market shares for the product with higher initial rating ($S_1^0 = 4$ and $S_2^0 = 2$).}
\label{theta0:paths}
\end{figure}
\end{center}
The social network used in this paper for simulations is a co-authorship network of scientists working on network theory and experiment~\cite{newman2006finding}. The network consists of 1589 nodes and 2743 edges and has the properties common for social networks such as power law degree distribution. 
Our simulations show that the market share of each option converges to some value, but there is not a unique limit point: different runs usually converge to different limits.  For instance, Figure~\ref{theta0:paths} shows three sample paths for the market share of option 1 when $\sigma = 1$ and $\theta = 0$ and initial star ratings of $S_1^0 = 4$ and $S_2^0=2$ stars respectively. We ran same simulations for other numbers of initial stars which were qualitatively the same. 

The multiplicity of limit points has been previously observed in settings with one type of social influence.  In particular, \cite{bhw05} shows that when objective evidence is weak (and thus decisions are only affected by social influence), then the limit point is not predetermined, or in other words, there are many potential limit points~\cite{bhw05}.  We observe that this also holds in a setting with two types of social influence (friends and the general public).  However, in contrast to the setting of Bendor et al., in our setting certain limit points are more likely than others.

In general, there is a range of likely limit points, which depends on the standard deviation $\sigma$ (of the normal distribution from which the ratings come from) and the threshold $\theta$.  We study the effect of $\sigma$ and $\theta$ on the distribution of limit points.

\begin{center}
\begin{figure}
\includegraphics[width=3in, height=1.5in]{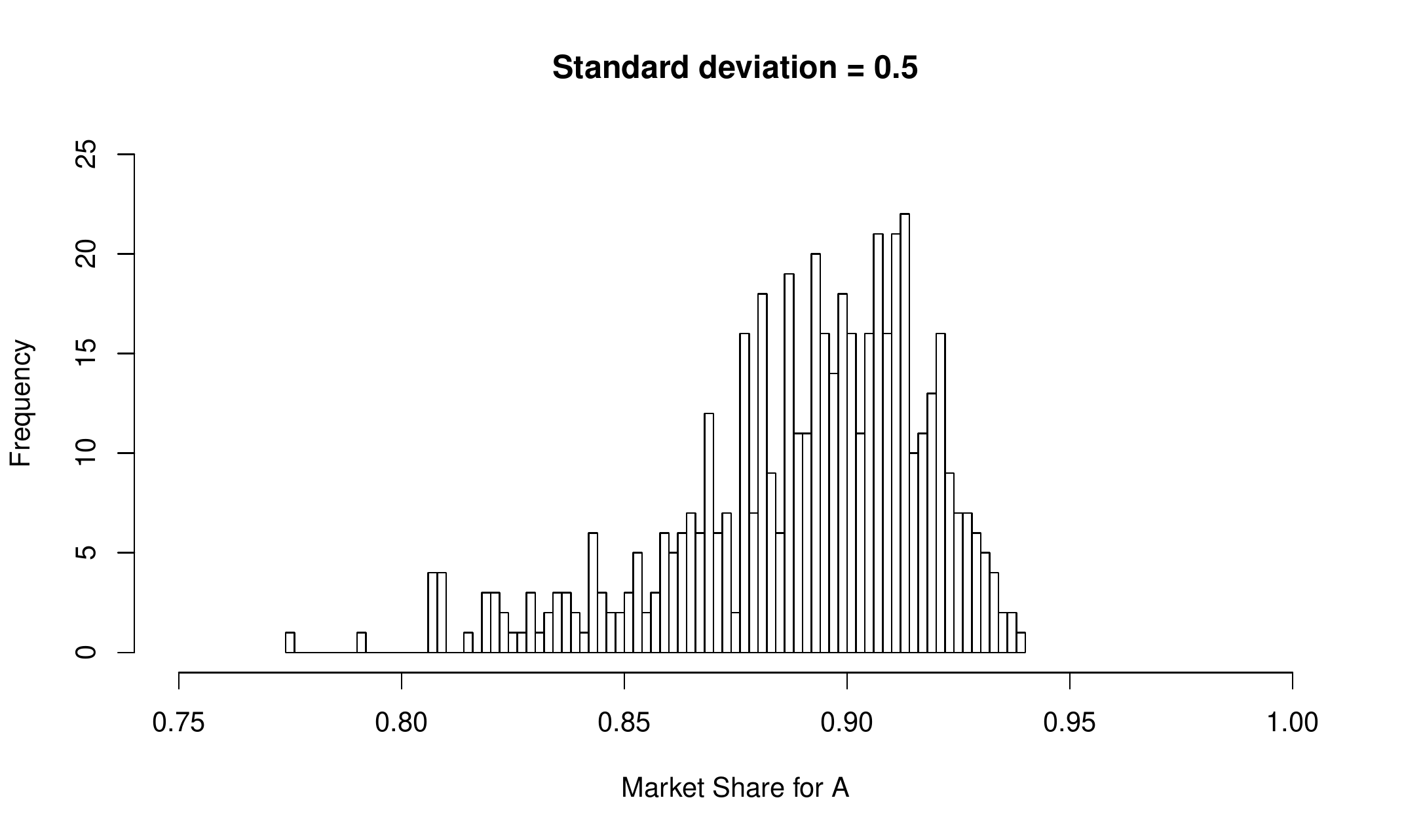}
\includegraphics[width=3in, height=1.5in]{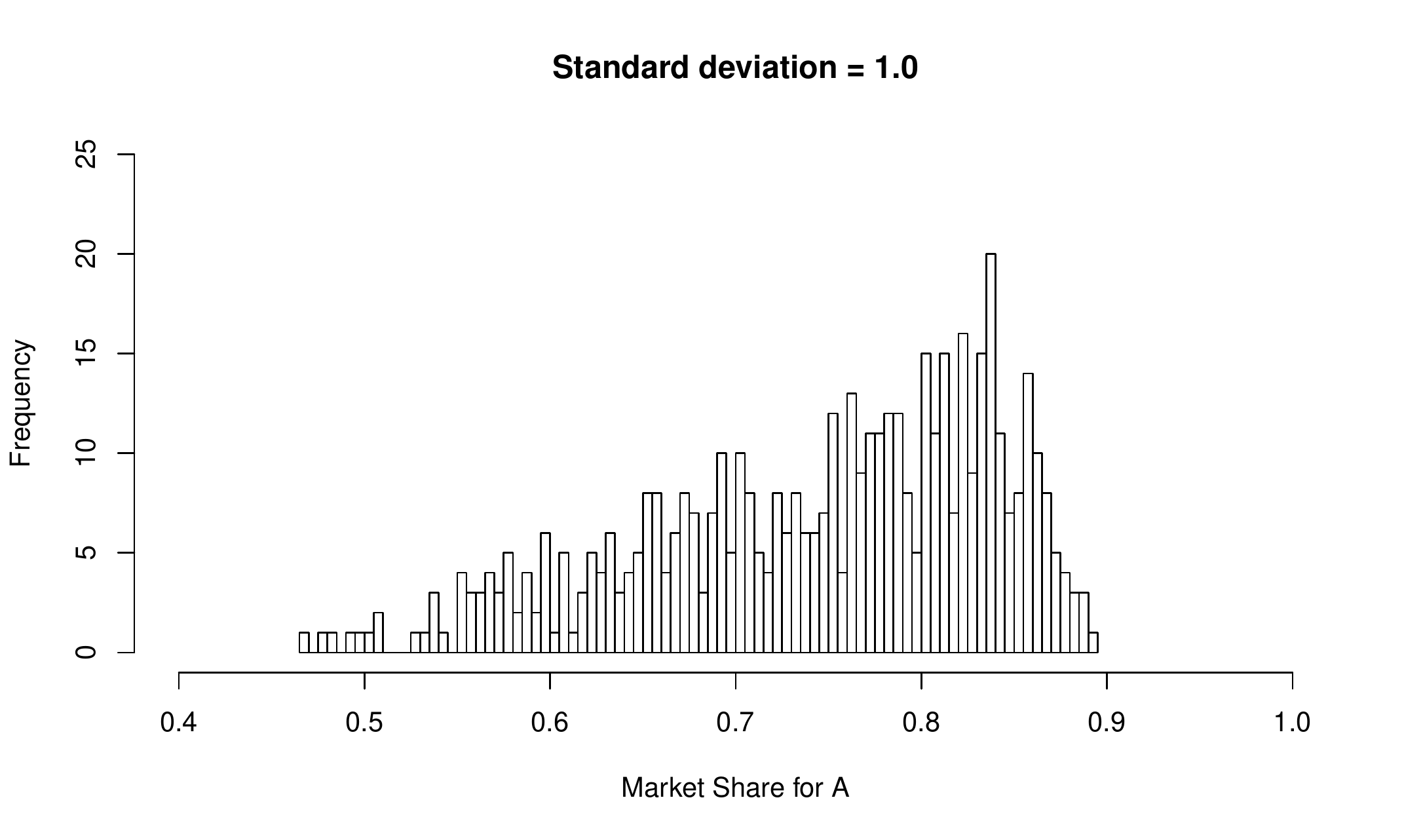}
\caption{Market share value frequencies over 500 runs ($\theta = 0$, and $S_1^0 =4 $ and $S_2^0 = 2$).}
\label{theta0:hist}
\end{figure}
\end{center}

Figure \ref{theta0:hist} shows the histograms for the frequency of different limit points when $\theta = 0$ for $\sigma = 0.5$ and $\sigma = 1$, $S_1^0 = 4$ and $S_2^0 = 2$ (based on 500 runs).  These histograms approximate the corresponding distributions of limit points.  We observe that the range of likely outcomes becomes narrower when the standard deviation is smaller. In other words, the variability of limit points increases when the variability of ratings increases.

\begin{center}
\begin{figure}
\includegraphics[width=3.5in, height=4.13in]{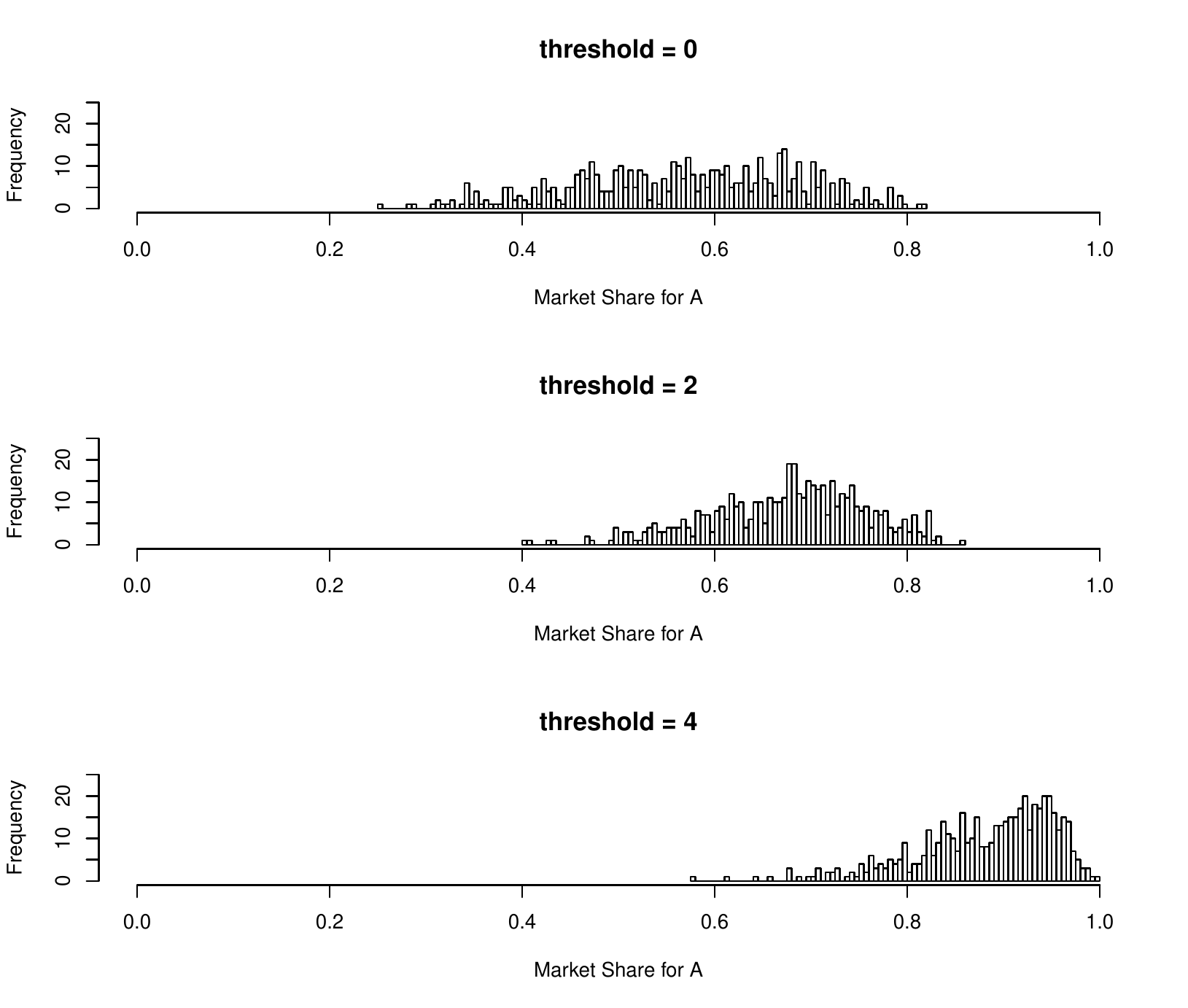}
\caption{Frequencies of market shares for the product with higher initial ratings over 500 runs ($\sigma = 1$, and $S_1^0 =3 $ and $S_2^0 =2 $).}
\label{thresholds}
\end{figure}
\end{center}

We next consider the effect of the threshold $\theta$ for a fixed $\sigma$.  
The histograms in Figure~\ref{thresholds} show the market share for option 1 for $\theta = 0,2,4$ when $\sigma = 1$ (based on 500 runs) with initial star ratings of $S_1^0=3$ and $S_2^0=2$.
We observe that as $\theta$ increases, the distribution of limit points increases (in the sense of first order stochastic dominance).  In particular, if $\theta_1 < \theta_2$, then the limit point distribution for $\theta_2$ stochastically dominates the distribution for $\theta_1$.  This implies that the product with the higher initial rating is more likely to have a larger market share for larger values of $\theta$.

Observe that as $\theta$ increases, individuals become more selective in terms of when to recommend an item to their friends.  When $\theta = 0$, an individual recommends any item she buys to her friends.  On the other extreme, when $\theta = 5$, an individual only recommends items that she is extremely satisfied with.  
Figure~\ref{thresholds} suggests that when individuals become more selective in their recommendations, then it becomes even more likely to have a large market share for the option that initially dominates (in terms of having more stars at time 0), or in other words, inequality increases.  

\section{Conclusion}
Our study of how online users make choices based on information from friend recommendations and ratings from the general public is important for a range of online applications in particular social search results
ranking, recommender systems, online advertisement placement, online social network
newsfeed rankings, and social shopping websites. When both friends' recommendations and ratings from the general public are available, our estimated
model can help the platform determine which option to display
or in what ranking to display the options for a given user.

Our results offer insights that can be useful in various online
domains. Specifically we found that (1) one additional star has a larger effect than one more friend recommendation, (2) negative opinions from friends are more influential than positive opinions, and (3) people show more random behavior in their choices when lower cost or risk is incurred. Our study of the dynamics of market share in the presence of these two forms of social influence reveals that the variability of outcomes increases when the variability of ratings increases, and inequality increases when individuals become more selective in their recommendations.

While this paper focuses on two sources of information, namely friends' opinions and ratings from the general public, our approach can also be applied to the study of how individuals trade off information from other sources, such as experts, celebrities, and the media.

\bibliographystyle{acm-sigchi}
\bibliography{sample}

\end{document}